\documentclass[12pt]{revtex4}
\usepackage{amsfonts}
\usepackage{graphicx, amsmath}
\usepackage{longtable}
\makeatletter

\newcommand{\Rmnum}[1]{\expandafter\@slowromancap\romannumeral #1@}
\makeatother
\begin{document}
\title[Short Title]{Dressed-state scheme for a fast CNOT gate}
\author{Jin-Lei Wu}
\author{Xin Ji\footnote{E-mail: jixin@ybu.edu.cn}}
\author{Shou Zhang}
\affiliation{Department of Physics, College of Science, Yanbian University, Yanji, Jilin 133002, People's Republic of China}

\begin{abstract}
A highly feasible dressed-state scheme, which greatly speeds up the adiabatic population transfer of a quantum system, is applied for implementing a fast CNOT gate in a cavity QED system. The cavity QED system consists of two identical five-level atoms which are respectively trapped in two optical cavities connected by a fiber. With the help of quantum Zeno dynamics, the CNOT gate is constructed by three steps with six pulses. Because the adiabatic condition is not necessary to be considered and all of the pulses are smoothly turned on and off, the scheme is fast and feasible in experiment. Numerical simulations indicate that the average fidelity for constructing the CNOT gate is quite high and the gate operation time is relatively short. Moreover, the effects on the fidelity of the atomic spontaneous emission and the photon leakages from the cavities are discussed by the master equation and the result shows the scheme is robust against the decoherence.
\end{abstract}
\maketitle
\section{Introduction}
Quantum logic gates are key elements of a quantum computer which possesses stronger computational
power and faster operational speed than a classical computer~\cite{LKG1998}. As we all know, any gate operation in quantum computation can be decomposed into a series of elementary one-qubit unitary gates and two-qubit conditional gates which are universal for quantum computation~\cite{DPD1995,AB1995}. Many schemes have been proposed to implement quantum gate operations in various physical systems, such as the ion-trap systems~\cite{JP1995}, cavity quantum electrodynamics~(QED) systems~\cite{ADA1995,XAS2007}, nuclear magnetic resonance systems~\cite{NI1997} and superconducting devices~\cite{CSS2003,CS2006}. The controlled-not~(CNOT) gate is an important two-qubit universal gate which can be used to construct any multiqubit gates combined with single-qubit gates. Up to the present, many schemes have been proposed for realizing a CNOT gate~\cite{ADB2000,JBT2004,NXS2006,DLT2009,XLS2009}. For example, Franson \emph{et al.} realized a CNOT gate through the quantum Zeno effect of two optical qubits by two-photon absorbtion in 2000~\cite{JBT2004}; Sangouard \emph{et al.} implemented a CNOT gate by adiabatic passage with an optical cavity in 2006~\cite{NXS2006}; Shao \emph{et al.} proposed two scheme to perform CNOT gates via quantum Zeno dynamics~(QZD) and  stimulated Raman adiabatic passage~(STIRAP), respectively, in 2009~\cite{XLS2009}.

As we can see from the references~\cite{ADB2000,JBT2004,NXS2006,DLT2009,XLS2009}, QZD and STIRAP are two techniques widely used to implement population transfer because of their robustness against decoherence in proper conditions. However, either QZD or STIRAP has its own unavoidable defects. On one hand, QZD is sensitive to the atomic spontaneous emission and variations in operation time. On the other hand, STIRAP usually requires a relatively long interaction time which will accumulate decoherence and destroy the desired dynamics. Therefore, many researchers have paid more attention to speeding up the adiabatic population transfer~\cite{MS20032008,MBB2009,XIA2010,ESS2013,SSX2011,XEJ2011,SXE2012,XJ2012,ADC2013,BGS20132014,YYQ2016,XQJ2016}. There are two methods, counterdiabatic driving~(transitionless quantum driving) and Lewis-Riesenfeld invariant, are widely used to speed up the adiabatic population transfer, and also many schemes have been proposed for constructing quantum gates based on the two methods~\cite{LXJ2014,YYQ2015,YLQ2015,YLC2015,YLX2015,JTD2015,ZYY2016}. However, unfortunately, the two methods unavoidably suffer from some experimental obstacles. On one hand, counterdiabatic driving requires either a direct coupling of the initial and target states~\cite{RLK1997,XIA2010,LE2014,SS2015} or a coupling not available in the original Hamiltonian~\cite{MMN2012}. On the other hand,  Lewis-Riesenfeld invariant leads to pulse schemes that either need an infinite energy gap to be
perfect~\cite{XJ2012}, or do not smoothly turn on or off~\cite{XJ2012,AA2014}. And thus, the two methods above are extremely challenging to implement in experiment.

A short time before, Baksic \emph{et~al.} proposed a new method to speed up adiabatic population transfer by using dressed states~\cite{AHA2016}. In Ref.~\cite{AHA2016}, the dressed states are skillfully defined to incorporate the nonadiabatic processes. The dressed-state scheme is more experimentally feasible than counterdiabatic driving and Lewis-Riesenfeld invariant, because it skillfully escapes from the experimental obstacles the latter two methods suffer from. Shortly after this, inspired by Ref.~\cite{AHA2016}, we apply creatively the dressed-state method to quantum state transfer and entanglement generation between two $\Lambda$-type atoms trapped in an optical cavity~\cite{JXS2016}. In this paper, we further apply the dressed-state method to a more complex quantum system, and achieve a more valuable and practical purpose, i.e., constructing a fast CNOT gate. The system is a cavity QED system which consists of two identical five-level atoms trapped respectively in two single-mode optical cavities connected by a fiber. We construct a fast CNOT gate by three steps, and two pulses are applied in each step, i.e., there are six pulses applied in the whole process.

This paper is organized as follows. In Sec. \ref{a}, we describe the physical model of the scheme and the general process for constructing the CNOT gate. In Sec. \ref{b}, we give the detailed dressed-state scheme for constructing the CNOT gate. In Sec. \ref{c}, the feasibility and the robustness for constructing the CNOT gate will be discussed by numerical simulations. The conclusion appears in Sec. \ref{d}.

\section{Physical model and general process}\label{a}

\begin{figure}[htb]
\begin{center}
\includegraphics[scale=0.8]{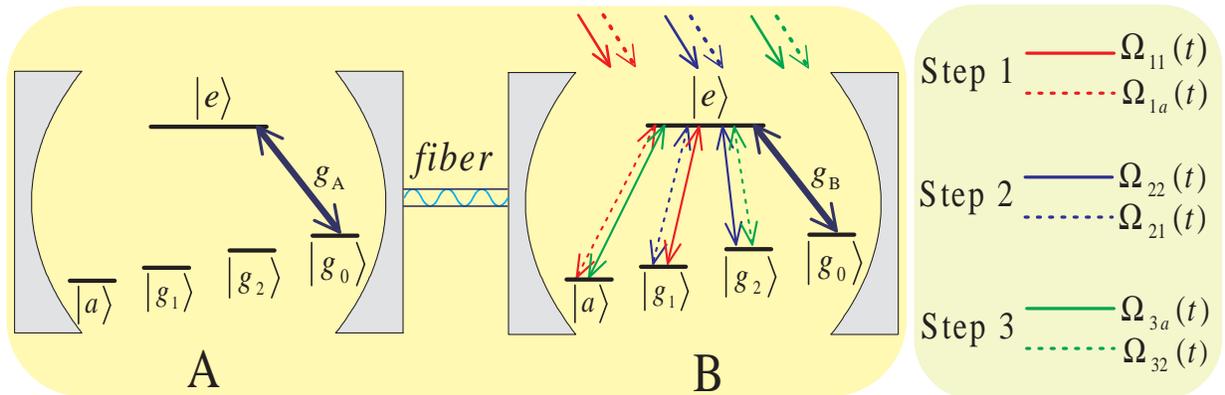}\\
\caption{(Color online) The diagrammatic sketch of cavity-atom combined system, atomic level configuration and related transitions.}\label{F1}
\end{center}
\end{figure}
The schematic setup for constructing a fast CNOT gate is shown in Fig.~\ref{F1}. Two identical five-level atoms A and B are trapped in two single-mode optical cavities, respectively. Each atom has an excited state $|e\rangle$ and four ground states $|a\rangle$, $|g_1\rangle$, $|g_2\rangle$ and $|g_0\rangle$. The atomic transitions $|e\rangle_{\rm A(B)}\leftrightarrow|g_0\rangle_{\rm A(B)}$ is resonantly coupled to the mode of the cavity \rm {A~(B)} with corresponding coupling constant $g_{\rm A(B)}$, and the transition $|e\rangle_{\rm B}\leftrightarrow|a\rangle_{\rm B}$, $|e\rangle_{\rm B}\leftrightarrow|g_1\rangle_{\rm B}$ and $|e\rangle_{\rm B}\leftrightarrow|g_2\rangle_{\rm B}$ are resonantly driven by classical fields with the time-dependent Rabi frequencies $\Omega_{\rm i}(t)$ shown at the right side in Fig.~\ref{F1} which will be described hereinafter in detail.

Now we show the general process of the construction of a fast CNOT gate. The quantum information is encoded in the states
\begin{eqnarray}\label{e1}
    &&|g_1g_1\rangle_{\rm AB}\equiv|00\rangle,\quad |g_1g_2\rangle_{\rm AB}\equiv|01\rangle, |g_0g_1\rangle_{\rm AB}\equiv|10\rangle,\quad |g_0g_2\rangle_{\rm AB}\equiv|11\rangle,
\end{eqnarray}
where $|g_ig_j\rangle_{\rm AB}~(i=0,1;j=1,2)$ donates the atoms \rm A and \rm B in the states $|g_i\rangle_{\rm A}$ and $|g_j\rangle_{\rm B}$, respectively. The system is initially in an arbitrary state
\begin{eqnarray}\label{e2}
    |\Psi_0\rangle&=&\sin\varepsilon\sin\beta|g_1g_1\rangle_{\rm AB}+\sin\varepsilon\cos\beta|g_1g_2\rangle_{\rm AB}+\cos\varepsilon\sin\beta|g_0g_1\rangle_{\rm AB}+\cos\varepsilon\cos\beta|g_0g_2\rangle_{\rm AB},\nonumber\\
\end{eqnarray}
where $\varepsilon$ and $\beta$ are two arbitrary complex angles. After a CNOT gate operation on the initial state $|\Psi_0\rangle$, the outcome state becomes
\begin{eqnarray}\label{e3}
    |\Psi\rangle&=&\sin\varepsilon\sin\beta|g_1g_1\rangle_{\rm AB}+\sin\varepsilon\cos\beta|g_1g_2\rangle_{\rm AB}
+\cos\varepsilon\sin\beta|g_0g_2\rangle_{\rm AB}+\cos\varepsilon\cos\beta|g_0g_1\rangle_{\rm AB}.\nonumber\\
\end{eqnarray}
Here, atom \rm A acts as the control qubit, and atom \rm B is the target qubit.

\begin{figure}[htb]
\begin{center}
\includegraphics[scale=0.8]{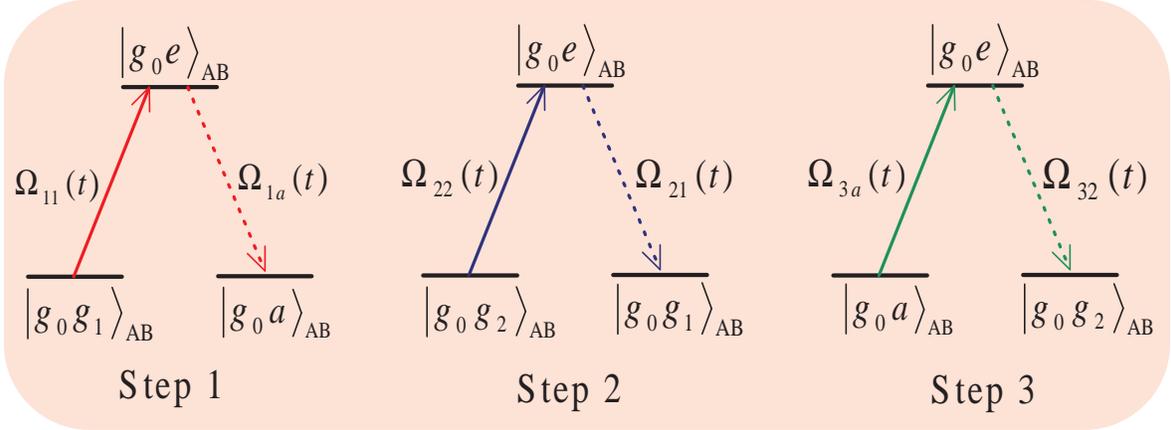}\\
\caption{(Color online) Schematic representation of the three steps for constructing a CNOT gate.}\label{F2}
\end{center}
\end{figure}

Only three steps are needed to achieve such a CNOT gate operation. Firstly, we implement the complete population transfer $|g_0g_1\rangle_{\rm AB}\rightarrow|g_0a\rangle_{\rm AB}$ by using the laser pulses interacting with atom \rm B to drive the atomic transitions $|e\rangle_{\rm B}\leftrightarrow|g_1\rangle_{\rm B}$ and $|e\rangle_{\rm B}\leftrightarrow|a\rangle_{\rm B}$ with the corresponding Rabi frequencies $\Omega_{11}(t)$ and $\Omega_{1a}(t)$, respectively. Then the initial state becomes
\begin{eqnarray}\label{e4}
    |\Psi_1\rangle&=&\sin\varepsilon\sin\beta|g_1g_1\rangle_{\rm AB}+\sin\varepsilon\cos\beta|g_1g_2\rangle_{\rm AB}
    +\cos\varepsilon\sin \beta|g_0a\rangle_{\rm AB}+\cos\varepsilon\cos\beta|g_0g_2\rangle_{\rm AB}.\nonumber\\
\end{eqnarray}
Secondly, we implement the complete population transfer $|g_0g_2\rangle_{\rm AB}\rightarrow|g_0g_1\rangle_{\rm AB}$ by using the laser pulses interacting with atom \rm B to drive the atomic transitions $|e\rangle_{\rm B}\leftrightarrow|g_2\rangle_{\rm B}$ and $|e\rangle_{\rm B}\leftrightarrow|g_1\rangle_{\rm B}$ with the corresponding Rabi frequencies $\Omega_{22}(t)$ and $\Omega_{21}(t)$, respectively. Then the state of the system becomes
\begin{eqnarray}\label{e5}
    |\Psi_2\rangle&=&\sin\varepsilon\sin\beta|g_1g_1\rangle_{\rm AB}+\sin\varepsilon\cos\beta|g_1g_2\rangle_{\rm AB}
    +\cos\varepsilon\sin \beta|g_0a\rangle_{\rm AB}+\cos\varepsilon\cos\beta|g_0g_1\rangle_{\rm AB}.\nonumber\\
\end{eqnarray}
Finally, we implement the complete population transfer $|g_0a\rangle_{\rm AB}\rightarrow|g_0g_2\rangle_{\rm AB}$ by using the laser pulses interacting with atom \rm B to drive the atomic transitions $|e\rangle_{\rm B}\leftrightarrow|a\rangle_{\rm B}$ and $|e\rangle_{\rm B}\leftrightarrow|g_2\rangle_{\rm B}$ with the corresponding Rabi frequencies $\Omega_{3a}(t)$ and $\Omega_{32}(t)$, respectively. And thus the state of the system becomes the expected state
\begin{eqnarray}\label{e6}
    |\Psi\rangle&=&\sin\varepsilon\sin\beta|g_1g_1\rangle_{\rm AB}+\sin\varepsilon\cos\beta|g_1g_2\rangle_{\rm AB}
    +\cos\varepsilon\sin \beta|g_0g_2\rangle_{\rm AB}+\cos\varepsilon\cos\beta|g_0g_1\rangle_{\rm AB}.\nonumber\\
\end{eqnarray}
The process above is equivalent to a CNOT gate operation, which indicates we can construct a CNOT gate by such three steps shown in Fig.~\ref{F2}.
\section{Dressed-state scheme for constructing the CNOT gate}\label{b}

In the following, we show the details for constructing the CNOT gate by using dressed states. For the first step, the time-dependent interaction Hamiltonian of the whole system is written as (setting $\hbar=1$)
\begin{eqnarray}\label{e7}
H_{1}(t)&=&H_{\rm al}(t)+H_{\rm acf}(t),\nonumber\\
H_{\rm al}(t)&=&\Omega_{11}(t)|e\rangle_{\rm B}\langle g_1|+\Omega_{1a}(t)|e\rangle_{\rm B}\langle a|+\mathrm{H.c.},\nonumber\\
H_{\rm acf}(t)&=&\sum_{k=\rm{A, B}}[g_{k}a_k|e\rangle_{k}\langle g_0|+\nu ba^\dag_k]+\mathrm{H.c.},
\end{eqnarray}
where $a_{\rm A(B)}$ is the annihilation operator of the mode of the cavity \rm {A(B)}, $b$ is the annihilation operator of the mode of the fiber mode, and $\nu$ is the coupling strength between the cavities' modes and the fiber mode. For simplicity, we assume $g_{\rm A}=g_{\rm B}=g$. Then with the initial state $|\Psi_0\rangle$ in Eq.~(\ref{e2}), dominated by the Hamiltonian~(\ref{e7}), the whole system evolves in the Hilbert space spanned by
\begin{eqnarray}\label{e8}
    |\phi_1\rangle&=&|g_1g_1\rangle_{\rm AB}|000\rangle,\quad
    |\phi_2\rangle=|g_1g_2\rangle_{\rm AB}|000\rangle,\nonumber\\
    |\phi_3\rangle&=&|g_0g_1\rangle_{\rm AB}|000\rangle,\quad
    |\phi_4\rangle=|g_0g_2\rangle_{\rm AB}|000\rangle,\nonumber\\
    |\phi_5\rangle&=&|g_1e\rangle_{\rm AB}|000\rangle,\quad
    |\phi_6\rangle=|g_0e\rangle_{\rm AB}|000\rangle,\nonumber\\
    |\phi_7\rangle&=&|g_1a\rangle_{\rm AB}|000\rangle,\quad
    |\phi_8\rangle=|g_1g_0\rangle_{\rm AB}|001\rangle,\nonumber\\
    |\phi_9\rangle&=&|g_0a\rangle_{\rm AB}|000\rangle,\quad
    |\phi_{10}\rangle=|g_0g_0\rangle_{\rm AB}|001\rangle,\nonumber\\
    |\phi_{11}\rangle&=&|g_1g_0\rangle_{\rm AB}|010\rangle,\quad
    |\phi_{12}\rangle=|g_0g_0\rangle_{\rm AB}|010\rangle,\nonumber\\
    |\phi_{13}\rangle&=&|g_1g_0\rangle_{\rm AB}|100\rangle,\quad
    |\phi_{14}\rangle=|g_0g_0\rangle_{\rm AB}|100\rangle,\nonumber\\
    |\phi_{15}\rangle&=&|eg_0\rangle_{\rm AB}|000\rangle,
\end{eqnarray}
in which the unsubscripted ket $|ijk\rangle~(i, j, k=0, 1)$ donates $i$, $j$ and $k$ photon in the cavity \rm A, fiber, and cavity \rm B, respectively. Because the initial state $|\Psi_0\rangle$ is the dark state of $H_{\rm acf}(t)$, i.e., $H_{\rm acf}(t)|\Psi_0\rangle=0$, by choosing the quantum Zeno limit condition $\Omega_{11}(t), \Omega_{1a}(t)\ll g,v$, the whole system can approximatively evolve in an invariant Zeno subspace consisting of dark states of $H_{\rm acf}(t)$~\cite{PS2002,PGS2009}
\begin{eqnarray}\label{e9}
    H_P=\Big\{|\phi_1\rangle, |\phi_2\rangle, |\phi_3\rangle, |\phi_4\rangle, |\phi_7\rangle, |\phi_9\rangle, |\phi_d\rangle  \Big\},
\end{eqnarray}
corresponding to the projections
\begin{equation}\label{e10}
    P^\alpha=|\alpha\rangle\langle\alpha|,\quad(|\alpha\rangle\in H_{P}).
\end{equation}
Here,
\begin{eqnarray}\label{e11}
    |\phi_d\rangle&=&\frac{1}{\sqrt{2\nu^2+g^2}}\Big(\nu|\phi_6\rangle-g|\phi_{12}\rangle+\nu|\phi_{15}\rangle\Big).
\end{eqnarray}
Therefore, by setting $v=g$, we can rewrite the system Hamiltonian as the following form~\cite{XLS2010}
\begin{eqnarray}\label{e12}
H(t)&\simeq&\sum_{\alpha}P^\alpha H_{\rm al}(t)P^\alpha~\nonumber\\
&=&\Omega_1(t)|\phi_3\rangle\langle\phi_d|+\Omega_2(t)|\phi_9\rangle\langle\phi_d|+\rm H.c.,
\end{eqnarray}
in which $\Omega_1(t)=\Omega_{11}(t)/\sqrt{3}$ and $\Omega_2(t)=\Omega_{1a}(t)/\sqrt{3}$.
Except $|g_0g_1\rangle_{\rm AB}$, obviously, $|g_1g_1\rangle_{\rm AB}$, $|g_1g_2\rangle_{\rm AB}$ and $|g_0g_2\rangle_{\rm AB}$ do not participate in the system evolution governed by the Hamiltonian~(\ref{e12}).

By choosing
\begin{equation}\label{e13}
    \Omega_1(t)=-\Omega(t)\sin\theta(t),\quad\Omega_2(t)=\Omega(t)\cos\theta(t),
\end{equation}
with $\Omega(t)=\sqrt{\Omega_1(t)^2+\Omega_2(t)^2}$ and $\theta(t)=\arctan(\Omega_1(t)/\Omega_2(t))$, we can easily obtain the time-dependent eigenstates of $H(t)$
\begin{eqnarray}\label{e14}
    |\varphi_d(t)\rangle&=&\cos\theta(t)|\phi_3\rangle+\sin\theta(t)|\phi_9\rangle,\nonumber\\
    |\varphi_{\pm}(t)\rangle&=&\frac{1}{\sqrt{2}}\Big(\sin\theta(t)|\phi_3\rangle\mp|\phi_d\rangle-\cos\theta(t)|\phi_9\rangle\Big),
\end{eqnarray}
with the eigenvalues $E_d=0$ and $E_{\pm}=\pm \Omega(t)$, respectively.

For convenience, we transform the time-dependent Hamiltonian~(\ref{e12}) to the time-independent eigen frame by the unitary operator $U(t)=\sum_{j=d,\pm}|\varphi_j\rangle\langle\varphi_j(t)|$. In the time-independent eigen frame the Hamiltonian~(\ref{e12}) becomes
\begin{eqnarray}\label{e15}
    H_{\rm ad}(t)=\Omega(t)M_z+\dot{\theta}(t)M_y,
\end{eqnarray}
where $M_z=|\varphi_+\rangle\langle\varphi_+|-|\varphi_-\rangle\langle\varphi_-|$ and $M_y=i(|\varphi_+\rangle+|\varphi_-\rangle)\langle\varphi_d|/\sqrt{2}+\rm H.c.$. In order to protect the system evolution from the second term of the Hamiltonian~(\ref{e15}) which leads to an imperfect population transfer, we introduce the modified Hamiltonian $H_{\rm mod}(t)=H(t)+H_c(t)$ to govern a perfect population transfer with the addition of a correction Hamiltonian $H_c(t)$. $H_c(t)$ can given by the general form
\begin{eqnarray}\label{e16}
    H_{c}(t)=U^\dag(t)(g_x(t)M_x+g_z(t)M_z)U(t),
\end{eqnarray}
where $M_x=(|\varphi_-\rangle-|\varphi_+\rangle)\langle\varphi_d|/\sqrt{2}+\rm H.c.$, $g_x(t)$ and $g_z(t)$ are two undetermined parameters. And thus the Hamiltonian~(\ref{e12}) becomes
\begin{eqnarray}\label{e17}
    H_{\rm mod}(t)&=&H(t)+H_c(t)\nonumber\\
    &=&\Omega'_1(t)|\phi_1\rangle\langle\phi_d|+\Omega'_2(t)|\phi_5\rangle\langle\phi_d|+\rm H.c.,
\end{eqnarray}
with the modified pulses
\begin{eqnarray}\label{e18}
    \Omega'_1(t)=g_x(t)\cos\theta(t)-[g_z(t)+\Omega(t)]\sin\theta(t),\nonumber\\
    \Omega'_2(t)=g_x(t)\sin\theta(t)+[g_z(t)+\Omega(t)]\cos\theta(t),
\end{eqnarray}
and then the Hamiltonian~(\ref{e7}) becomes
\begin{eqnarray}\label{e19}
H^{\rm mod}_{1}(t)&=&\Omega'_{11}(t)|e\rangle_{\rm B}\langle g_1|+\Omega'_{1a}(t)|e\rangle_{\rm B}\langle a|+\sum_{k=\rm{A, B}}[g_{k}a_k|e\rangle_{k}\langle g_0|+\nu ba^\dag_k]+\mathrm{H.c.},
\end{eqnarray}
with $\Omega'_{11}(t)=\sqrt{3}\Omega'_1(t)$ and $\Omega'_{1a}(t)=\sqrt{3}\Omega'_2(t)$.

With reference to Ref.~\cite{AHA2016}, we introduce a set of dressed states $|\tilde{\varphi}_{\pm,d}(t)\rangle$ by the unitary operation $|\tilde{\varphi}_{\pm,d}(t)\rangle=V(t)|\varphi_{\pm,d}\rangle$. We choose the unitary operator~\cite{JXS2016}
\begin{eqnarray}\label{e20}
V(t)=\exp[i\mu(t)M_x],
\end{eqnarray}
with an Euler angle $\mu(t)$. After transforming the modified Hamiltonian~(\ref{e17}) to the frame defined by $V(t)$, the Hamiltonian~(\ref{e17}) becomes
\begin{eqnarray}\label{e21}
H_{\rm new}(t)&=&V H_{\rm ad}(t)V^\dag+VU H_c(t)U^\dag V^\dag+i\frac{d V}{dt}V^\dag\nonumber\\
&=&\eta(t)\Big(|\tilde{\varphi}_+(t)\rangle\langle\tilde{\varphi}_{+}(t)|-|\tilde{\varphi}_{-}(t)\rangle\langle\tilde{\varphi}_{-}(t)|\Big)\nonumber\\
&&+\xi(t)\Big[\Big(|\tilde{\varphi}_+(t)\rangle-|\tilde{\varphi}_{-}(t)\rangle\Big)\langle\tilde{\varphi}_{d}(t)|+\rm H.c.\Big],
\end{eqnarray}
with the time-dependent parameters
\begin{eqnarray}\label{e22}
\eta(t)&=&[g_z(t)+\Omega(t)]\cos\mu(t)-\dot{\theta}(t)\sin\mu(t)\nonumber\\
\xi(t)&=&\frac{1}{\sqrt{2}}\Big\{i[g_z(t)+\Omega(t)]\sin\mu(t)+i\dot{\theta}(t)\cos\mu(t)+[\dot{\mu}(t)-g_x(t)]\Big\}.
\end{eqnarray}
After a simple calculation, we can choose
\begin{eqnarray}\label{e23}
g_x(t)=\dot{\mu}(t),\quad g_z(t)=-\Omega(t)-\frac{\dot{\theta}(t)}{\tan\mu(t)},
\end{eqnarray}
to remove the second term of the Hamiltonian~(\ref{e21}). Then back to the time-dependent adiabatic frame, the dark dressed state, which corresponds the zero eigenvalue of $H_{\rm new}(t)$, is written as
\begin{eqnarray}\label{e24}
|\tilde{\varphi}_{d}(t)\rangle&=&\cos\mu(t)\Big[\cos\theta(t)|\phi_3\rangle+\sin\theta(t)|\phi_9\rangle\Big]+i\sin\mu(t)|\phi_d\rangle.
\end{eqnarray}
Obviously, if the parameters satisfy $\theta(t_{i})=0$, $\theta(t_{1f})=\pi/2$ and $\mu(t_{i})=\mu(t_{1f})=0$, where $t_{{i}({1f})}$ is the initial~(final) time of the first step for constructing the CNOT gate, the desired population transfer $|\phi_3\rangle\rightarrow|\phi_9\rangle$ will be achieved by the system evolution along the dark dressed state $|\tilde{\varphi}_{d}(t)\rangle$.

Based on the process above, we have achieved the transfer $|g_0g_1\rangle_{\rm AB}\rightarrow|g_0a\rangle_{\rm AB}$, but the states $|g_1g_1\rangle_{\rm AB}$, $|g_1g_2\rangle_{\rm AB}$ and $|g_0g_2\rangle_{\rm AB}$ remain unchanged. Therefore, the first step is achieved for constructing the CNOT gate. Besides, the evolution process is not necessarily slow and there is no direct coupling between the initial and target state, as long as a set of suitable dressed states are chosen.

Similar to the first step, the modified Hamiltonian of the second step for constructing the CNOT gate is written as
\begin{eqnarray}\label{e25}
H^{\rm mod}_{2}(t)&=&\Omega'_{22}(t)|e\rangle_{\rm B}\langle g_2|+\Omega'_{21}(t)|e\rangle_{\rm B}\langle g_1|+\sum_{k=\rm{A, B}}[g_{k}a_k|e\rangle_{k}\langle g_0|+\nu ba^\dag_k]+\mathrm{H.c.},
\end{eqnarray}
in which
\begin{eqnarray}\label{e26}
\Omega'_{22}(t)&=&\sqrt{3}\Big\{g_x(t-t_{1f})\cos\theta(t-t_{1f})-[g_z(t-t_{1f})+\Omega(t-t_{1f})]\sin\theta(t-t_{1f})\Big\},\nonumber\\
\Omega'_{21}(t)&=&\sqrt{3}\Big\{g_x(t-t_{1f})\sin\theta(t-t_{1f})+[g_z(t-t_{1f})+\Omega(t-t_{1f})]\cos\theta(t-t_{1f})\Big\}.
\end{eqnarray}
If the parameters satisfy $\theta(t_{1f})=0$, $\theta(t_{2f})=\pi/2$ and $\mu(t_{1f})=\mu(t_{2f})=0$, where $t_{{1f}({2f})}$ is the initial~(final) time of the second step for constructing the CNOT gate, the desired transfer $|g_0g_2\rangle_{\rm AB}\rightarrow|g_0g_1\rangle_{\rm AB}$ will be achieved.

Similarly, the modified Hamiltonian of the third step for constructing the CNOT gate is written as
\begin{eqnarray}\label{e27}
H^{\rm mod}_{3}(t)&=&\Omega'_{3a}(t)|e\rangle_{\rm B}\langle a|+\Omega'_{32}(t)|e\rangle_{\rm B}\langle g_2|+\sum_{k=\rm{A, B}}[g_{k}a_k|e\rangle_{k}\langle g_0|+\nu ba^\dag_k]+\mathrm{H.c.},
\end{eqnarray}
in which
\begin{eqnarray}\label{e28}
\Omega'_{3a}(t)&=&\sqrt{3}\Big\{g_x(t-t_{2f})\cos\theta(t-t_{2f})-[g_z(t-t_{2f})+\Omega(t-t_{2f})]\sin\theta(t-t_{2f})\Big\},\nonumber\\
\Omega'_{32}(t)&=&\sqrt{3}\Big\{g_x(t-t_{2f})\sin\theta(t-t_{2f})+[g_z(t-t_{2f})+\Omega(t-t_{2f})]\cos\theta(t-t_{2f})\Big\}.
\end{eqnarray}
If the parameters satisfy $\theta(t_{2f})=0$, $\theta(t_{3f})=\pi/2$ and $\mu(t_{2f})=\mu(t_{3f})=0$, where $t_{{2f}({3f})}$ is the initial~(final) time of the third step for constructing the CNOT gate, the desired transfer $|g_0a\rangle_{\rm AB}\rightarrow|g_0g_2\rangle_{\rm AB}$ will be achieved. By now, the transform $|\Psi_0\rangle\rightarrow|\Psi\rangle$ is achieved and we implement the CNOT gate successfully.
\section{Numerical simulations}\label{c}
In this section, the feasibility and the robustness for constructing the CNOT gate will be discussed by numerical simulations.
\begin{figure}[htb]
\begin{center}
\includegraphics[scale=0.7]{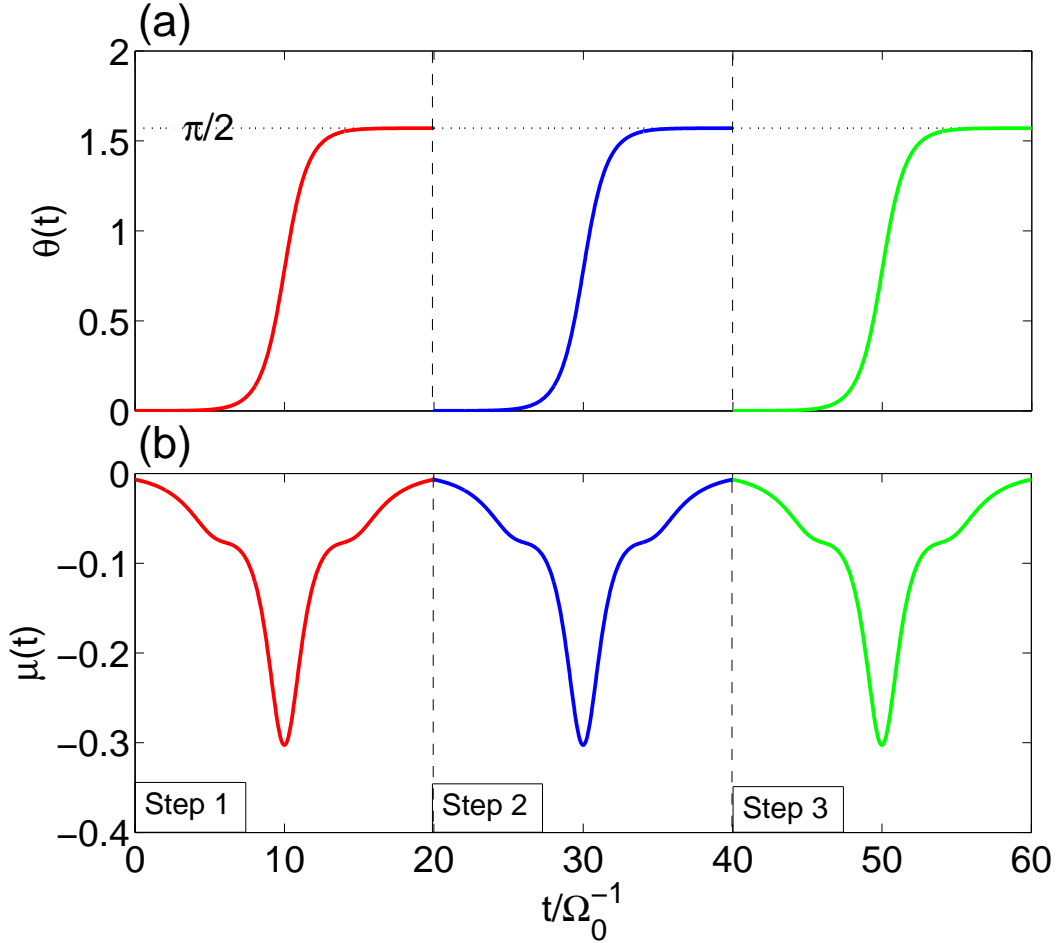}\\
\caption{(Color online) $\theta(t)$ and $\mu(t)$ versus $t/\Omega_0^{-1}$. The parameters used here are \{$t_f=20/\Omega_0$, $\tau=t_0=0.1t_f$\}.}\label{F3}
\end{center}
\end{figure}

First of all, $\Omega_1(t)$ and $\Omega_2(t)$ can be chosen as the Gaussian pulses~\cite{KHB1998,NTB2001}
\begin{eqnarray}\label{e29}
\Omega_1(t)&=&\Omega_0\exp[-(t-t_f/2-t_0)^2/\tau^2],\nonumber\\
\Omega_2(t)&=&\Omega_0\exp[-(t-t_f/2+t_0)^2/\tau^2],
\end{eqnarray}
in which the related parameters are chosen as $t_f=20/\Omega_0$ and $\tau=t_0=0.1t_f$. Besides, there exist the relations $t_{1f}=t_f$, $t_{2f}=2t_f$ and $t_{3f}=3t_f$. $\mu(t)$ is defined by
\begin{eqnarray}\label{e30}
\mu(t)=-\arctan\Big(\frac{\dot{\theta}(t)}{g(t)/\tau+\Omega(t)}\Big),
\end{eqnarray}
where $g(t)=\mathrm{sech}(t/\tau)$ is chosen to regularize $\mu(t)$ such that it can meet the condition $\mu(t_i)=\mu(t_f)=0$ and make $\sin^2\mu(t)$, which is the population of $|\phi_d\rangle$~(see Eq.~(\ref{e24})), as small as possible. Then, in order to check whether or not the parameters we choose are suitable, we plot $\theta(t)$ and $\mu(t)$ versus $t/\Omega_0^{-1}$ in Fig.~\ref{F3}. Without a doubt, figure \ref{F3} shows that the chosen parameters are in accord with all of the boundary conditions \{$\theta(t_{i})=0$, $\theta(t_{1f})=\pi/2$, $\mu(t_{i})=\mu(t_{1f})=0$\}, \{$\theta(t_{1f})=0$, $\theta(t_{2f})=\pi/2$, $\mu(t_{1f})=\mu(t_{2f})=0$\} and \{$\theta(t_{2f})=0$, $\theta(t_{3f})=\pi/2$, $\mu(t_{2f})=\mu(t_{3f})=0$\} for constructing the CNOT gate.
\begin{figure}[htb]
\begin{center}
\includegraphics[scale=0.7]{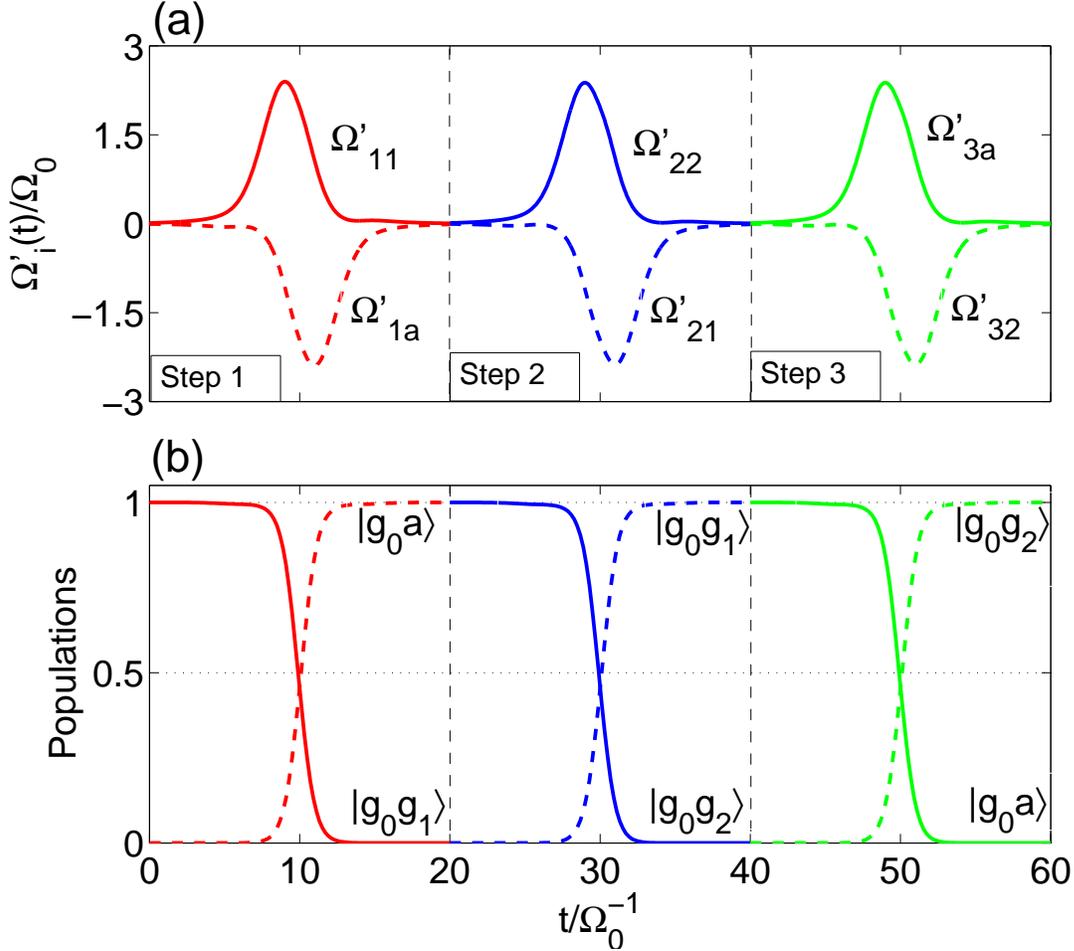}\\
\caption{(Color online) (a)~The six time-dependent modified pulses versus $t/\Omega_0^{-1}$; (b)~the time-dependent population conversions of the three steps for constructing the CNOT gate versus $t/\Omega_0^{-1}$. The parameters used here are \{$t_f=20/\Omega_0$, $\tau=t_0=0.1t_f$, $g=v=10\Omega_0$\}.}\label{F4}
\end{center}
\end{figure}

Based on the relations \{$\Omega'_{11}(t)=\sqrt{3}\Omega'_1(t)$, $\Omega'_{1a}(t)=\sqrt{3}\Omega'_2(t)$, Eq.~(\ref{e26}), Eq.~(\ref{e28})\}, we can determine the six time-dipendent modified pulses $\Omega'_{11}(t)$, $\Omega'_{1a}(t)$, $\Omega'_{22}(t)$, $\Omega'_{21}(t)$, $\Omega'_{3a}(t)$ and $\Omega'_{32}(t)$. In Fig.~\ref{F4}(a), we plot the six modified pulses versus $t/\Omega_0^{-1}$. And in Fig.~\ref{F4}(b), we plot the time-dependent population conversions of the three steps for constructing the CNOT gate, respectively. Here, for the Zeno limit condition $\Omega_{\rm i}(t)\ll g, v$, we choose $g=v=10\Omega_0$. From Fig.~\ref{F4}(a), we can clearly find that the six modified pulses are completely available Gaussian pulses and smoothly turned on and off. In addition, Fig.~\ref{F4}(b) shows that all of the expected population conversions are achieved for constructing the CNOT gate.

\begin{figure}[htb]
\begin{center}
\includegraphics[scale=0.7]{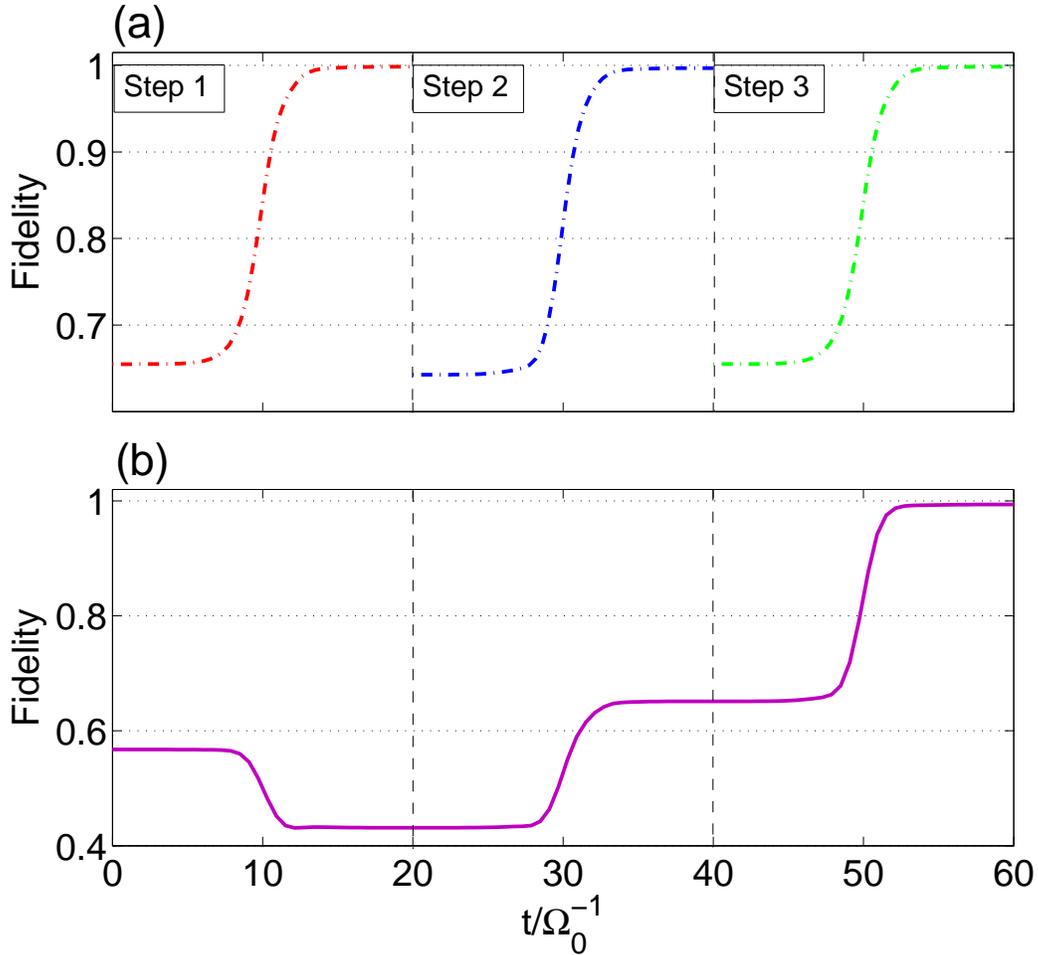}\\
\caption{(Color online) (a)~the average fidelity of each step versus $t/\Omega_0^{-1}$; (b)~the average fidelity of the whole gate process versus $t/\Omega_0^{-1}$. The parameters used here are same as those in Fig.~\ref{F4}.}\label{F5}
\end{center}
\end{figure}

\begin{figure}[htb]
\begin{center}
\includegraphics[scale=0.7]{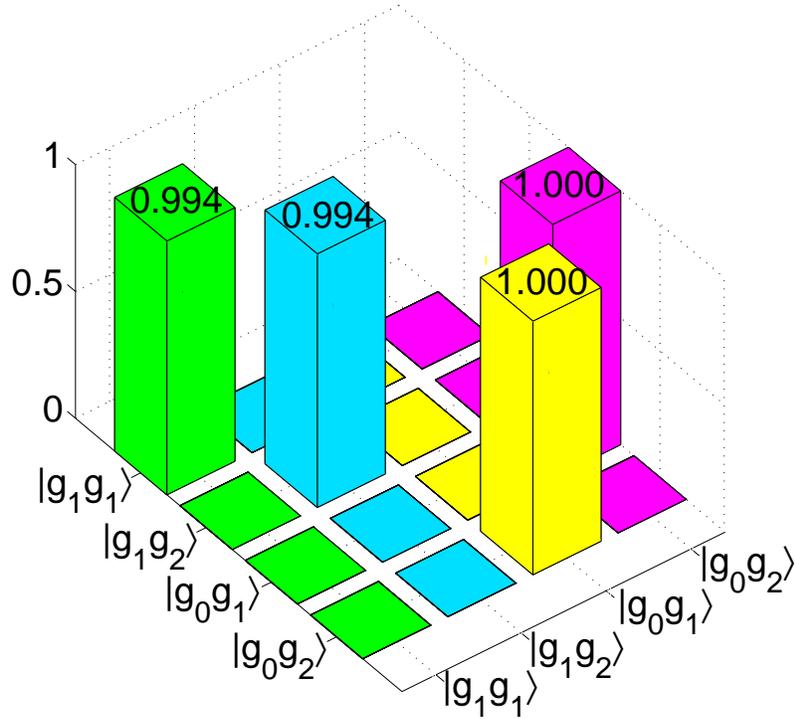}\\
\caption{(Color online) The truth table of the CNOT gate. The parameters used here are same as those in Fig.~\ref{F4}.}\label{F6}
\end{center}
\end{figure}

Next, we show the average fidelity of each step in Fig.~\ref{F5}(a) and that of the whole gate process in Fig.~\ref{F5}(b), respectively. The average fidelity is defined by
\begin{eqnarray}\label{e31}
F=\frac{1}{(2\pi)^2}\int^{2\pi}_{0}\int^{2\pi}_{0}|\langle\Psi_{ideal}|\Psi_{t}\rangle|^2d\varepsilon d\beta.
\end{eqnarray}
Here, $|\Psi_{ideal}\rangle=|\Psi_{1}\rangle$ for the step 1; $|\Psi_{ideal}\rangle=|\Psi_{2}\rangle$ for the step 2; $|\Psi_{ideal}\rangle=|\Psi\rangle$ for the step 3 and the whole gate process. $|\Psi(t)\rangle$ is the state of the system governed by the Hamiltonian~(\ref{e19}) for the step 1, Hamiltonian~(\ref{e25}) for the step 2, or Hamiltonian~(\ref{e17}) for the step 3. As shown in Fig.~\ref{F5}(a), the average fidelity of each step is near unit at each finial time. Correspondingly, in Fig.~\ref{F5}(b), the average fidelity of the whole gate process is $F=0.994$ at $t=60/\Omega_0$, which indicates our scheme is highly feasible within a very short gate operation time. As a clearer illustration, in Fig.~\ref{F6}, we give a truth table of the CNOT gate constructed by the above dressed-state scheme, which also indicates our scheme is highly feasible.

In the following, we take the effect of decoherence on the dressed-state scheme into account. The whole system is dominated by the master equation
\begin{eqnarray}\label{e32}
\dot{\rho}(t)&=&-i[H^{\rm mod}(t),\rho(t)]\nonumber\\
\cr&&-\sum_{j=\rm A,B}\sum_{k=a,g_1,g_2,g_0}\frac{\gamma_j}{2}\Big[\sigma_{e_j,e_j}\rho-2\sigma_{k_j,e_j}\rho\sigma_{e_j,k_j}+\rho\sigma_{e_j,e_j}\Big]\nonumber\\
\cr&&-\sum_{j=\rm A,B}\frac{\kappa_j}{2}\Big[a^{\dag}_ja_j\rho-2a_j\rho a^{\dag}_j+\rho a^{\dag}_ja_j\Big]\nonumber\\
\cr&&-\frac{\kappa_f}{2}\Big[b^{\dag}b\rho-2b\rho b^{\dag}+\rho b^{\dag}b\Big],
\end{eqnarray}
where $H^{\rm mod}(t)=H^{\rm mod}_i(t)$ for the $i$th step~($i=1,2,3$). $\gamma_{\rm A(B)}$ is the spontaneous emission rate of atom \rm A(B) from the excited state $|e\rangle_{\rm A(B)}$ to the ground state $|k\rangle_{\rm A(B)}~(k=a,g_1,g_2,g_0)$; $\kappa_{\rm A(B)}$ denotes the photon leakage rate from the cavity \rm A(B); $\sigma_{e_j,k_j}=|e\rangle_{j}\langle k|$. For simplicity, we assume $\gamma_{\rm A}=\gamma_{\rm B}=\gamma/4$ and $\kappa_{\rm A}=\kappa_{\rm B}=\kappa_{f}=\kappa$. Then the average fidelity is rewritten as
\begin{eqnarray}\label{e33}
F=\frac{1}{(2\pi)^2}\int^{2\pi}_{0}\int^{2\pi}_{0}|\langle\Psi_{ideal}|\rho(t)|\Psi_{ideal}\rangle|d\varepsilon d\beta.
\end{eqnarray}
\begin{figure}[htb]
\begin{center}
\includegraphics[scale=0.7]{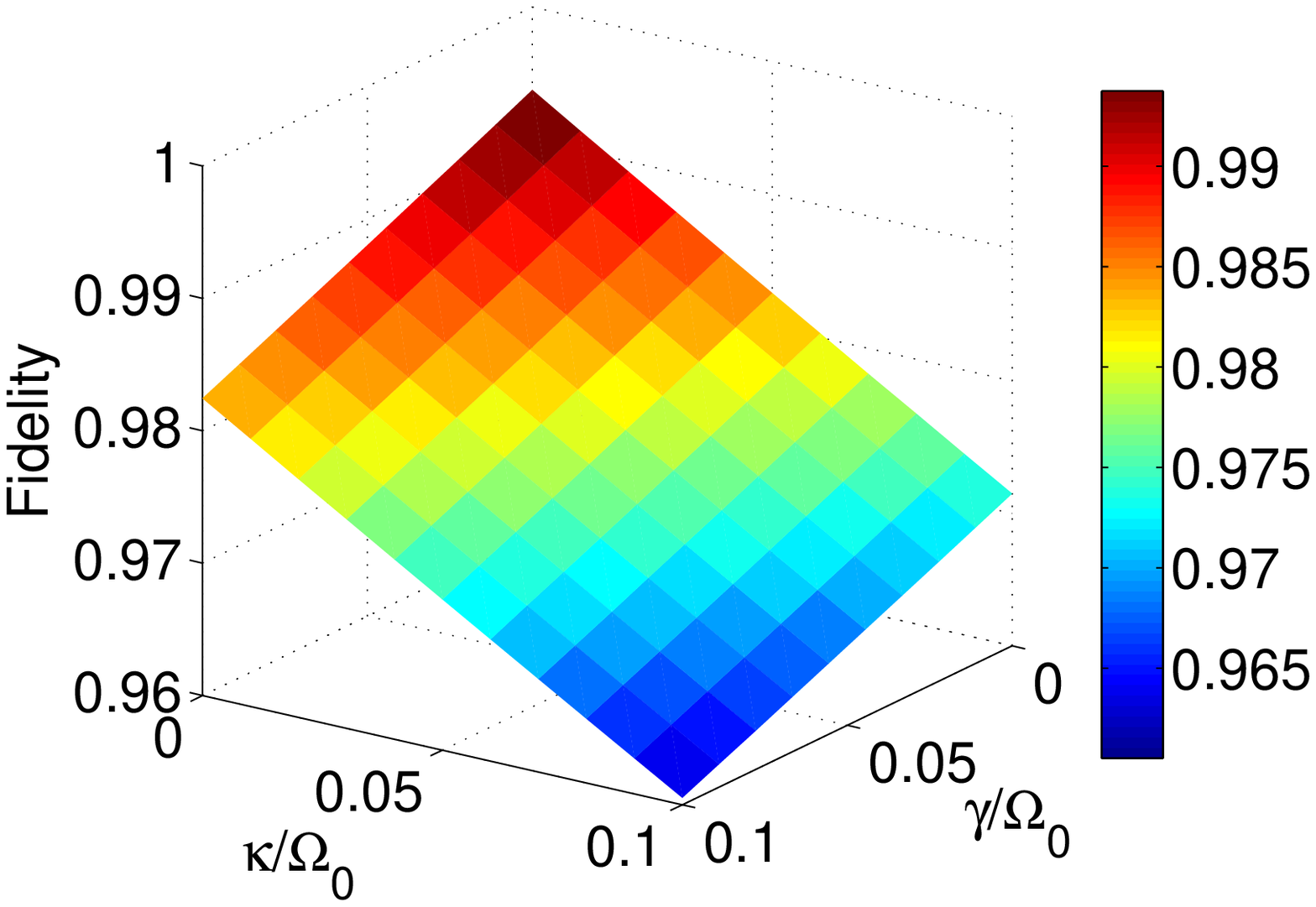}\\
\caption{(Color online) The fidelity as a function of $\gamma/\Omega_{0}$ and $\kappa/\Omega_{0}$. The parameters used here are the same as those in Fig.~\ref{F4}.}\label{F7}
\end{center}
\end{figure}

Based on the above master equation, we plot the effect of decoherence on the average fidelity of the whole process for constructing the CNOT gate in Fig.~\ref{F7}. As we can see from Fig.~\ref{F7}, we learn that the influence of the photon leakages from the cavities on the fidelity is obviously greater than that of  the atomic spontaneous emissions. It is not difficult to understand it. As we all know, QZD can effectively restrain the influence of the photon leakages from the cavities if the Zeno limit condition is met. But in our scheme, the amplitude of the six modified pulses is near $2.5\Omega_0$~(see Fig.~\ref{F4}~(a)) and the chosen parameters $g=v=10\Omega_0$ do not satisfy the Zeno limit condition $\Omega'_{\rm i}\ll g, v$ strictly. Even so, however, the fidelity of the whole process for constructing the CNOT gate are over 0.96 even when $\kappa=\gamma=0.1\Omega_0$. Therefore, the dressed-state scheme for constructing the CNOT gate is robust against the decoherence induced by the atomic spontaneous emissions and the photon leakages from the cavities.

\section{Conclusion}\label{d}
In conclusion, we have proposed a highly feasible dressed-state scheme for constructing a fast CNOT gate in a cavity QED system. Different from counterdiabatic driving and Lewis-Riesenfeld invariants, there is no a direct coupling of the target state and the initial state appearing in the Hamiltonian and all of the modified pulses used in the scheme are smoothly turned on and off, which ensure the feasibility of the scheme in practice. During the whole process for constructing the CNOT gate, the adiabatic condition need not to be satisfied, and thus the gate operation is fast. The results of the numerical simulations indicate that the construction of the CNOT gate is robust against the decoherence induced by the atomic spontaneous emissions and the photon leakages from the cavities. Considering the current experimental conditions, by using cesium atoms and a set of predicted cavity QED parameters $(g,\kappa,\gamma)/2\pi=(750,3.3,2.62)$ MHz~\cite{STK2005}, the CNOT gate can be constructed with a fidelity $F=0.99$. Hence, even in the presence of decoherence, the dressed-state scheme for constructing a fast CNOT gate is also highly feasible. In a word, by using dressed-state scheme, a fast, feasible and robust CNOT gate is constructed. Furthermore, we believe that other more complex quantum gates can be constructed in appropriate quantum systems by using dressed-state scheme.
\\
\begin{center}
{\bf{ACKNOWLEDGMENT}}
\end{center}
This work was supported by the National Natural Science Foundation of China under Grants No. 11464046 and No. 61465013.

\end{document}